\documentclass[11pt]{article}

\usepackage{caption}
\usepackage{amssymb,latexsym, amsthm}
\usepackage[latin1]{inputenc}
\usepackage{graphics,epsfig,graphicx}
\usepackage{amsmath}
\usepackage{amsfonts}
\usepackage{dsfont}
\usepackage[english]{babel}
\usepackage{algorithm,algorithmic,hyperref}
\usepackage{verbatim}
\usepackage{booktabs} 
\usepackage{pstricks}
\usepackage{pst-node,pst-tree,pst-eps,xspace}
\IfFileExists{pst-beta.tex}
            {\input{pst-beta.tex}}
           {\RequirePackage{pst-tree}}

\numberwithin{equation}{section}
\numberwithin{equation}{section}
\newtheorem{theorem}{Theorem}
\newtheorem{lemma}{Lemma}
\newtheorem{proposition}{Proposition}
\newtheorem{corollary}{Corollary}

\theoremstyle{definition}
\newtheorem{ex}{Example}
\newtheorem{definition}{Definition}
\newcommand{\comm}[1]{#1}

\newcommand{\seqnum}[1]{\href{http://oeis.org/#1}{\tt{#1}}}
\begin{document}
\title{Exhaustive generation for permutations avoiding a (colored) regular sets of patterns}

\author{Phan Thuan Do \footnote{Department of Computer Science, 
Hanoi University of Science and Technology, 01 Dai Co Viet, Hanoi, Vietnam,
Email: thuandp@soict.hust.edu.vn} \and Thi Thu Huong Tran  \footnote{Vietnamese-German University, Le Lai street, Hoa Phu ward, Thu Dau Mot city, Binh Duong, Vietnam, Email: huong.ttt@vgu.edu.vn} \and
Vincent Vajnovszki \footnote{LE2I, Universit\'e de Bourgogne Franche-Comt\'e, B.P. 47 870, 21078 Dijon-Cedex, France, Email: vvajnov@u-bourgogne.fr}}

\maketitle

\begin{abstract} 
Despite the fact that the field of pattern avoiding permutations has been skyrocketing over the last two decades,
there are very few exhaustive generating algorithms for such classes of permutations.
In this paper we introduce the notions of regular and 
colored regular set of forbidden patterns, which are particular cases of 
right-justified sets of forbidden patterns.
We show the (colored) regularity of several  sets of forbidden patterns 
(some of them involving variable length patterns) and 
we derive a general framework for the efficient generation of permutations avoiding them. The obtained generating algorithms are based on succession functions,
a notion which is a byproduct of the ECO method introduced in the context of enumeration and random generation of combinatorial objects
by Barcucci et al. in 1999, and developed later by 
Bacchelli et al. in 2004, for instance.
For some classes of permutations falling under our general framework, 
the corresponding counting sequences are classical in combinatorics, such as Pell, Fibonacci, Catalan, Schr\"oder and
binomial transform of Padovan sequence.
\end{abstract}

\section{Introduction}

Given a class of combinatorial objects, it is a common problem
to list exhaustively (with no repetitions nor omissions) all the objects with a given size in the class.
Exhaustive generation can be used to test hypotheses about a class of objects,
to support a conjecture or find counterexamples, to analyze or prove programs, etc,
and often an exhaustive generating algorithm exhibits new properties 
of the class under consideration. Two books, that of F. Ruskey \cite{Rus17} and more recently that of D. Knuth 
\cite{Knu2011} are entirely devoted 
to the exhaustive generation of combinatorial objects.

When generating combinatorial objects, the time complexity of a generating
algorithm is crucial since the cardinality of a
class is, in general, an exponential function of the size of the generated objects. 
If a generating algorithm produces combinatorial objects
so that only a constant amount of computation is done between successive
objects, in an amortized sense, then one says that it runs in 
constant amortized time (or CAT) \cite{Rus17}.

The field of pattern avoiding permutations has been showing an increasing
interest in the last two decades. However little has been done so far on the 
exhaustive generation of such classes of permutations.
A very powerful way to define, enumerate and construct recursively  
such permutation classes is the ECO method~\cite{BBGP04,BDPP99}. 
This is a general recursive description of combinatorial classes which explains how an 
object of a given size can be reached uniquely from an object of smaller size. 
More specifically, the  ECO method specifies through a succession function 
how many objects of larger size can be obtained from an object of a given size.

In this paper, which is an enhanced version of the conference version \cite{DTV},
we give a general framework for the efficient (that is, CAT) exhaustive generation of some classes of pattern avoiding
permutations. The classes under consideration are defined by means of regular and colored regular 
set of forbidden patterns, both defined in this paper. They cover a wide range of pattern avoiding 
permutations and many of them are enumerated by classical integer sequences.
A main ingredient for our generating algorithms is the succession function corresponding
to the set of forbidden patterns.

\medskip

After the presentation of some basic definitions at the end of this section, the remainder of the paper is structured 
as follows. In Section \ref{R-j} we recall the notion of right-justified set of forbidden patterns,
originally introduced in \cite{Do08, DFMV08}, and we give a characterization of such sets of patterns.
Section \ref{S:3} is devoted to regular sets of forbidden patterns, a subclass of right-justified ones
where the succession functions are computationally efficient. We show that a particular set of forbidden patterns
involving two variable length patterns is regular, and some instances of it yield
known counting sequences for the corresponding pattern avoiding permutations.
This notion is further refined in Section \ref{S:4} to colored regular sets of forbidden patterns, 
and we show the colored-regularness of some sets of forbidden patterns (one of them involving a 
variable length pattern) and, as previously, some known enumerating sequences are obtained.
Notice that the idea of color labeling in the ECO generating context has previously been mentioned in ~\cite{BBP06,BD06}.
In the last section we present a general framework for the efficient exhaustive generation 
for permutations avoiding a regular or a colored regular set of forbidden patterns.
Finally, in Appendix, we give a list of regular and colored regular sets of forbidden patterns together with their succession functions. Each of these classes can be exhaustively generated in constant amortized time by our algorithms.



\subsubsection*{Pattern avoiding permutations}

We denote by $S_n$ the set of permutations on $\{1,2,\dots,n\}$, $n\geq 0$, 
and the empty permutation $\epsilon$ 
is the unique permutation in $S_0$.
We use the one-line notation: for $\pi\in S_n$ we write $\pi=\pi(1)\pi(2)\ldots\pi(n)$, where $\pi(i)$ is the image of  $i$ by $\pi$, and $n$ is said the {\it length} of $\pi$.
A permutation $\sigma$ is {\it contained} in another permutation $\pi$ if $\pi$ has a 
(not necessarily contiguous) subsequence whose terms are order isomorphic to (i.e., have same relative ordering as) 
$\sigma$. In this context $\sigma$ is called a {\it pattern}, and if $\sigma$ is not contained in $\pi$
we say that $\pi$ {\it avoids} $\sigma$. 
For example the permutation $461532\in S_6$ contains the pattern $312$ because the sequence $413$
(among others) is ordered in the same way as $312$, whereas $24531\in S_5$ avoids $312$. 

For a set of patterns $P$ we say that a permutation avoids $P$ (or, it is 
$P$-avoiding), if it avoids each
pattern in $P$, and in this context $P$ is called {\it set of forbidden patterns}.
We denote by $S_n(P)$
the set of length $n$ permutations avoiding $P$:
$$
S_n(P)=
\cap_{\sigma\in P}\{\pi\in S_n: \pi \text{ avoids } \ \sigma\},$$
and $S(P)=\cup_{n\geq 0}S_n(P)$.
The set $S(P)$ is a downset in the permutation pattern involvement order, that is, $\pi$ belongs to 
$S(P)$ whenever  $\pi$ occurs as a pattern in a permutation in $S(P)$.
See S. Kitaev's seminal book \cite{Kitaev2011} for an extensive presentation of 
pattern avoidance in permutations.


\section{Right-justified forbidden patterns} 
\label{R-j}

Here we introduce the right-justified forbidden patterns,
a particular class of forbidden patterns defined in \cite{Do08, DFMV08},
and we give a characterization of them. In the next two sections we refine this notion to
(colored) regular patterns.

Let $P$ be a set of forbidden patterns.
Each permutation in $S_n(P)$, $n\geq 1$, can be obtained from a unique one in $S_{n-1}(P)$ by inserting the entry $n$ into the appropriate position.
Informally, $P$ is said to be right-justified if, for any $n\geq 1$, it 
satisfies the following property:
if the insertion of $n$ into the position $i$
of $\alpha\in S_{n-1}(P)$ yields a permutation in $S_n(P)$, then so does the 
insertion of $n$ into any position to the right of $i$ in $\alpha$.




Let $\alpha$ be a length $n$ permutation. We denote by $\alpha^{\rightarrow}$ (resp. $\alpha^{\leftarrow}$) the permutation obtained from $\alpha$ by moving its largest entry $n$ to the right (resp. left) one position;
and $\alpha^{\rightarrow}$ (resp. $\alpha^{\leftarrow}$) is defined only if $\alpha(n)\ne n$ (resp. $\alpha(1)\ne n$). For instance, $3142^{\rightarrow}=3124$ and $3142^{\leftarrow}=3412$. 

\begin{definition}
\label{def3}
The set of forbidden patterns $P$ is said to be 
\emph{right-justified} if 
$\alpha \in S(P)$ implies $\alpha^{\rightarrow} \in S(P)$. 
\end{definition}
In other words, $P$ is right-justified if for any length $n$ permutation avoiding $P$ by moving $n$ to the right we still obtain a permutation avoiding $P$. 

\begin{ex} $ $
\begin{itemize}
\item $P =  \{132\}$ is not right-justified since, for instance, $3412 \in S_4(132)$ but 
$3412^{\rightarrow}=3142\notin S_4(132)$.
It is easily seen that $P=\{312\}$ is right-justified, and in general, a singleton set of forbidden patterns $P=\{\tau\}$ with $\tau\in S_k$ is right-justified if and only if $\tau(1) = k$.
\item $P=\{312, 123\}$ is not right-justified since, for instance,  $132\in S_3(P)$ but 
$132^{\rightarrow}=123\notin S_3(P)$. 
By contrast, $P=\{312,132\}$ is right-justified and Theorem \ref{P:rjp2} below gives a characterization of right-justified forbidden patterns. 
\end{itemize}
\end{ex}

The right-justifiedness of a set of forbidden pattern is a prerequisite for its regularity,
a notion introduced in \cite{DFMV08} and presented in the next section. Many permutation patterns considered in the literature are regular, although their right-justifiedness was considered only implicitly or even omitted. 
The next theorem gives a characterization of right-justified patterns. 


\begin{theorem}\label{P:rjp2}
The set $P$ of (possibly different lengths) forbidden patterns is right-justified if and only if
for any $\tau\in P$, if $\tau^{\leftarrow}$ exists, then it contains a pattern in $P$. 
\end{theorem}

\proof
\comm{
Assume that $P$ is a set of right-justified forbidden patterns. Let $\tau\in P$
and suppose that $\tau^{\leftarrow}$ exists and it avoids $P$. 
Since $P$ is right-justified, we have $(\tau^\leftarrow)^\rightarrow=\tau$ avoids $P$, which is a contradiction. 

\noindent
Conversely, let $P$ be a set of forbidden patterns such that $\tau^{\leftarrow}$ contains a pattern in $P$ for any $\tau\in P$, whenever $\tau^{\leftarrow}$ exists.
We need to prove that $\alpha^{\rightarrow}\in S_n(P)$
for any $\alpha \in S_n(P)$ whenever $\alpha^{\rightarrow}$ exists.
Let $\alpha\in S_n(P)$ and let $i<n$ be such that $\alpha(i)=n$. Thus, by definition
$$\alpha^{\rightarrow}=\alpha(1)\dots \alpha(i-1)\alpha(i+1)\, n\, \dots \alpha(n),$$ 
and let suppose that $u=u_1u_2\ldots u_k$ is an occurrence of the length $k$
pattern $\tau\in P$ in the permutation $\alpha^{\rightarrow}$. We distinguish two cases.

If $u$ does not contain simultaneously entries $\alpha(i+1)$ and $n$, then $u$ is 
a subsequence of $\alpha$ too, so $\alpha$ contains $\tau$, which is a contradiction.

Otherwise, let $u'$ be the sequence obtained from $u$ by transposing $\alpha(i+1)$ and $n$,
that is, $u'$ has the form $u_1u_2\ldots \,n\, \alpha(i+1)\ldots u_k$. 
Clearly, $u'$ is an occurrence of $\tau^{\leftarrow}$ in $\alpha$, so 
$\alpha$ contains $\tau^{\leftarrow}$, which in turn contains a pattern in $P$, and thus
$\alpha\notin S_n(P)$.
This yields again to a contradiction. 
}
\endproof

In particular, if the patterns of $P$ have the same length, then 
we have the next simpler characterization.

\begin{corollary}
\label{prop1} Let $P$ be a set of forbidden patterns of same length. 
Then $P$ is right-justified if and only if for each $\tau\in P$ we have $\tau^{\leftarrow} \in P$,
whenever $\tau^{\leftarrow}$ exists. 
\end{corollary}

\noindent
Below are several examples of sets of forbidden patterns
whose right-justifiedness follows directly from Theorem \ref{P:rjp2}.
Some of these patterns will be considered in more details in the next two sections.

\begin{ex}$ $
\label{ex_d}
\begin{enumerate}
\item $P=\{321, 231\}$, and $|S_n(P)|=2^{n-1}$ for $n\geq 1$ 
      (sequence \seqnum{A000079} in OEIS \cite{OEIS}).
\item \label{bisection}
       $P=\{321, 3412\}$ and $P=\{312,2431\}$, and for both $|S_n(P)|$ gives the bisection of
Fibonacci sequence (\seqnum{A001519} in OEIS \cite{OEIS}).
\item $P=\{2134, 2143, 2413, 4213\}$, and $|S_n(P)|$ gives the central binomial coefficients 
      $\binom{2n-2}{n-1}$ (sequence \seqnum{A000984} in OEIS \cite{OEIS}).

\item $P=\{312,321,23\ldots(p+1)1\}$, $p\geq 2$, and $|S_n(P)|$ gives the sequence of $p$-generalized Fibonacci
      numbers. When $p=2$, $P$ becomes $\{312,321,231\}$ and $|S_n(P)|$ gives
      the sequence of Fibonacci numbers (\seqnum{A000045} in OEIS \cite{OEIS}), see for instance
      \cite{BBP06,BD06}.
\item \label{case_p}
      $P=\{312,2431,(p+1)p\ldots 21\}$, $p\geq 2$. When $p=3$, $P$ becomes $\{312,2431,4321\}$ and 
      $|S_n(P)|$ is the binomial transform of Padovan sequence (\seqnum{A034943} in OEIS
     \cite{OEIS}), see \cite{BD06}.
\item $P=\{321,p(p+1)12\ldots (p-1),(m+1)12\ldots m\}$, $p,m\geq 2$. In particular,
\begin{itemize} \label{case_p_m}
\item when $p=2$, $P=\{231,321,(m+1)12\ldots m\}$, and $|S_n(P)|$ gives
      again the generalized Fibonacci sequence. In particular, if $m=2$, then 
      $P=\{231,312,321\}$ and as above $|S_n(P)|$ gives the sequence of Fibonacci numbers \cite{BBP06};

\item when $p=m=3$, 
      $P=\{321, 3412, 4123\}$, and $|S_n(P)|$ gives
      the sequence of Pell numbers (\seqnum{A215928} in OEIS \cite{OEIS}), 
      see for instance \cite{BBP06, Gui95}.
\end{itemize}
\end{enumerate}
\end{ex}

\section{Regular patterns}\label{S:3}

Let $P$ be a set of right-justified forbidden patterns and $\alpha\in S_n(P)$.
As we have seen in the previous section, if the insertion of $(n+1)$ into the $i$th position 
of $\alpha$ yields a permutation in $S_{n+1}(P)$, then so does the insertion of $(n+1)$ into any position to the right of $i$.
In order to formalize this phenomenon, we define 
a {\it site} of a permutation as a position between two of its entries, and before the first and after the last entries. Sites are numbered from right to left, and so the rightmost site, that which follows the last entry of the permutation, is numbered by one. And by convention, the length zero
permutation $\epsilon$ has one site (numbered by one).

For $\alpha\in S_n$, we denote by $\alpha^{\downarrow i}$ the permutation obtained from $\alpha$ by inserting $(n+1)$ into its $i$th site. 
For a set of forbidden patterns $P$ and $\alpha\in S_n(P)$,
a site $i$ of $\alpha$ is called {\it active} (with respect to $P$) 
if $\alpha^{\downarrow i} \in S_{n+1}(P)$.
For a set $P$ of right-justified forbidden patterns active sites 
of a permutation $\alpha\in S(P)$ form an interval of integers beginning by $1$,
and if $i$ is an active site of $\alpha\in S(P)$ we denote by 
$\chi_P(i,\alpha)$ the number of active sites of $\alpha^{\downarrow i}$. 
It follows that if  $\beta=\alpha^{\downarrow i}\in S_{n+1}(P)$ for some 
$\alpha\in S_n(P)$ and an active site $i$ of $\alpha$, then $\beta^{\downarrow j}\in S_{n+2}(P)$ 
if and only if $j$ belongs to the interval $[1,\chi_P(i,\alpha)]$;
and any permutation in $S_{n+2}(P)$ can be obtained uniquely in this way from appropriate $\alpha$,
$i$ and $j$.

It can happen that $\chi_P(i,\alpha)$ does not depend on $\alpha$ but only on 
the number of active sites of $\alpha$, and we have the next definition.

\begin{definition}[Regular pattern/succession function]
\label{def_reg_p}
 A set $P$ of right-justified forbidden patterns
is called \emph{regular} if for any $\alpha\in  S_n(P)$, $n\geq 0$, 
\begin{itemize}
\item $\alpha$ has its first site active, and
\item if $\alpha$ has $k$ active sites, then for any $i$, $1\leq i\leq k$,
the number $\chi_P(i,\alpha)$ of active sites of $\alpha^{\downarrow i}$
does not depend on $\alpha$ but solely on $i$ and on 
$k$. In this case, $\chi_P(i,\alpha)$ is denoted by $\chi_P(i,k)$, and 
$$
\chi_P:\{(i,k)\,|\,k\in\mathbb{N}^+, 1\leq i\leq k\}\to\mathbb{N}^+
$$
is called a \emph{succession function}. 
\end{itemize} 
\end{definition}

Even we will not use explicitly later, it is worth to mention that for regular sets of forbidden patterns $P$ characterized by the succession function $\chi_P$, the set of productions
$$
\{(k) \rightsquigarrow  (\chi_P(1,k))(\chi_P(2,k)) \ldots (\chi_P(k,k))
\}_{k\geq 1}
$$
is called succession rule corresponding to $P$.
These productions are the core of ECO-method introduced in \cite{BBGP04} 
and were widely used afterwards in
more general contexts, as the enumeration or (random and exhausive) generation 
of combinatorial objects.

The following theorem gives the succession function for the set of right-justified
forbidden patterns in Example \ref{ex_d}.\ref{case_p_m}, and thus shows that it is a set of 
regular such patterns.

\begin{theorem} \label{Th.3pm}
The succession function corresponding to $P=\{321,p(p+1)12\ldots (p-1),(m+1)12\ldots m\}$,
$p,m\geq 2$, is defined by:

        \[\chi_P(i,k)=\left\{\begin{array}{lll}
            k+1, & \quad \mbox{if $i = 1$ and $k<m$} \\
            m, & \quad \mbox{if $i = 1$ and $k = m$} \\
            i, & \quad \mbox{if $1<i<p$} \\
            p-1, & \quad \mbox{otherwise.}
        \end{array}   \right. \]
    \end{theorem} 
\proof
\comm{


Let $\alpha=\alpha(1)\alpha(2)\dots\alpha(n)$ be a permutation in $S_n(P)$ with $k$ active sites. 
We consider the number of active sites of $\alpha^{\downarrow i}$ for each $i$, $1\leq i\leq k$, and 
since $P$ is right-justified, this number is precisely the maximum 
(i.e., the number of the leftmost) site 
$j$ of $\alpha^{\downarrow i}$ such that $(\alpha^{\downarrow i})^{\downarrow j}$ avoids $P$. 
We rewrite $\alpha^{\downarrow i}$ as 
$$\alpha^{\downarrow i}= \alpha(1)\alpha(2)\dots \alpha(n-i+1)(n+1)\alpha(n-i+2)\dots \alpha(n).$$ 

Since $\alpha^{\downarrow k}$ avoids $321$, the length $(k-1)$ suffix $\alpha(n-k+2) \dots \alpha(n)$
of $\alpha$ is increasing, otherwise the suffix $(n+1)\alpha(n-k+2)\ldots \alpha(n)$ of 
$\alpha^{\downarrow k}$ contains the forbidden pattern $321$. 
Moreover, since $\alpha^{\downarrow k}$ avoids $(m+1)12\dots m$, we have $k\leq m$, otherwise the 
same suffix $(n+1)\alpha(n-k+2)\dots\alpha(n)$ of $\alpha^{\downarrow k}$ contains the forbidden pattern $(m+1)12\dots m$. 
\medskip

\noindent 
If $i=1$ and $k<m$, then the permutation $(\alpha^{\downarrow 1})^{\downarrow (k+1)}$ does not contain
the pattern $(m+1)12\ldots m$ because $k<m$.
In addition, $(\alpha^{\downarrow 1})^{\downarrow (k+1)}$ contains neither $321$ nor 
$p(p+1)12\ldots (p-1)$, otherwise $\alpha^{\downarrow k}$ contains the same pattern.

\noindent
Moreover, $(\alpha^{\downarrow 1})^{\downarrow (k+2)}$ contains at least one of the patterns in $P$,
otherwise $\alpha^{\downarrow (k+1)}\in S_{n+1}(P)$. It follows that, in this case,
$\chi_P(i,k)=k+1$.
\medskip

\noindent 
If $i=1$ and $k=m$, reasoning in the same manner, the insertion of $(n+2)$ into $\alpha^{\downarrow 1}$
in any site less than or equal to $m$  does not produce patterns in $P$, 
but the insertion of $(n+2)$ into the $(m+1)$st site produces the pattern  $(m+1)12\ldots m$
as a suffix of $(\alpha^{\downarrow 1})^{\downarrow (m+1)}$.
It follows that, in this case, $\chi_P(i,k)=m$.
\medskip

\noindent 
If $1<i<p$, the insertion of $(n+2)$ into the $(i+1)$st site of $\alpha^{\downarrow i}$
produces the pattern $321$, whereas the insertion of $(n+2)$ into any site
less than or equal to  $i$  of $\alpha^{\downarrow i}$ does not produce patterns in $P$, thus
in this case, $\chi_P(i,k)=i$.
\medskip

\noindent
If $i\geq p$, from $i\leq k$ it follows that $p\leq k$, and so the insertion of $(n+2)$ into
the $p$th site of $\alpha^{\downarrow i}$ produces the pattern $p(p+1)12\ldots (p-1)$, but
the insertion of $(n+2)$ into any site of $\alpha^{\downarrow i}$ less than $p$ does not 
produce patterns in $P$. Thus in this case $\chi_P(i,k)=p-1$.
}
\endproof

Notice that, as mentioned in Example \ref{ex_d}.\ref{case_p_m}, particular instances of 
$m$ and $p$ give classical set of patterns:
when $m=p=3$, $P$ becomes  $\{321, 3421,4123\}$ investigated in \cite{BBP06, Gui95}
and the corresponding $P$-avoiding permutations are counted by Pell numbers
(\seqnum{A000129} in OEIS \cite{OEIS});
and when $p=2$ the obtained $P$-avoiding permutations are counted by the generalized Fibonacci numbers.

We give succession functions for some sets of regular forbidden patterns in Table \ref{table2} in Appendix.

\section{Colored regular patterns}\label{S:4}

Not surprisingly, any right-justified set of forbidden patterns $P$ is 
not necessarily  regular: 
it can happen that $\alpha$ and $\beta$ are $P$-avoiding permutations having
the same number of active sites, but the insertion of the next largest value 
into the $i$th active site of both $\alpha$ and $\beta$ yields permutations
with different numbers of active sites.
This section is devoted to the investigation of a particular class of such right-justified forbidden patterns
that we call, following Barcucci \emph{et al.} \cite{BBP06}, colored regular 
forbidden patterns. For such forbidden patterns we develop corresponding succession functions
and explicit them for two sets of forbidden patterns: 
that in Example \ref{ex_d}.\ref{case_p} and the second one in Example \ref{ex_d}.\ref{bisection}.

Let $\alpha$ be a permutation with $k$ active sites belonging to $S(P)$, 
with $P$ a right-justified set of forbidden patterns, and let $i$ be an active site of  $\alpha$, $1\leq i \leq k$. 
Suppose that it exists a procedure coloring by integer values the permutations in $S(P)$, so that: 
{\it (i)} the number of active sites of $\alpha^{\downarrow i}$ does not depend on $\alpha$ but only on the three parameters $i$, $k$ and the color $c$ of $\alpha$; and {\it (ii)}
the color of $\alpha^{\downarrow i}$ in turn, depends only on $i$, $k$ and $c$.
In this case we extend the function $\chi_P$ in the previous section  
so that it transforms the triple $(i,k,c)$ into a pair of integers:
the number of active sites and the color of $\alpha^{\downarrow i}$. In order to anchor the recursivity we set 
the color of the length zero permutation $\epsilon$ to $0$, and we have the next definition.

\begin{definition}[c-regular pattern]
A set $P$ of right-justified forbidden patterns is called {\em colored regular} 
{\it (c-regular} for short) if for any $\alpha\in S(P)$,
\begin{itemize}
    \item $\alpha$ has its first site active, and
    \item if $\alpha$ has $k$ active sites and color $c$, then
    for any $i$, $1\leq i\leq k$
         \begin{itemize}
          \item the number of active sites of $\alpha^{\downarrow i}$
                depends only on $i$, $k$ and $c$, 
                and we denote this number by $\mu_P(i,k,c)$,
          \item the color of $\alpha^{\downarrow i}$ depends as above only on $i$, $k$ and
                $c$, and we denote this color by $\nu_P(i,k,c)$.
         \end{itemize}
In this case, the succession function $\chi_P=(\mu_P,\nu_P)$ becomes:
$$
\chi_P=(\mu_P,\nu_P):
\{(i,k,c)\,|\,k\in\mathbb{N}^+, 1\leq i\leq k,c\in C\}\to\mathbb{N}^{+}\times C,
$$
where $C\subset \mathbb{N}$ is the set of colors.

\end{itemize}
\end{definition}

Notice that regular patterns are particular c-regular patterns, where the set of colors collapses  
to $\{0\}$.

Now we consider the set of right-justified forbidden patterns $P$
in Example \ref{ex_d}.\ref{case_p}, and the next theorem shows that $P$ is 
a c-regular set of forbidden patterns by giving explicitly its colored succession function. We postpone its proof after giving some technical results. 

\begin{theorem}\label{T:genpad}
The colored succession function for the set of forbidden patterns $P=\{312,2431,(p+1)p\ldots 21\}$,
$p\geq 2$, is $\chi_P(i,k,c)=(\mu_P(i,k,c),\nu_P(i,k,c))$, with set of colors $\{0,1\}$ and:
\begin{itemize}
\item[] $\mu_P(i,k,c)=\left\{\begin{array}{ll}
            i+1 & \mbox{if $i=1$ or ($i=k$ and $c=0$ and $k<p$)} \\
            i & \mbox{otherwise,}
        \end{array}   \right. $
\end{itemize}
\noindent and
\begin{itemize}
\item[] $\nu_P(i,k,c)=\left\{\begin{array}{ll}
            0 & \mbox{if $i=1$ or ($i=k$ and $c=0$)} \\
            1 & \mbox{otherwise.}
        \end{array}   \right.$
\end{itemize}
    \end{theorem}
\begin{lemma}\label{L:act1}$ $
\begin{enumerate}
\item If $\alpha=\alpha(1)\alpha(2)\ldots\alpha(n)$ is a length 
$n$ permutation avoiding $312$, and $i$ is such that $\alpha(i)=n$, then 
the suffix $\alpha(i)\alpha(i+1)\ldots\alpha(n)$ of $\alpha$ is decreasing.
\item \label{second_p}
Let $P$ be the set of forbidden patterns in Theorem \ref{T:genpad}.
If $\alpha\in S_n(P)$ has $k$ active sites, and 
$\beta=\alpha^{\downarrow i}\in S_{n+1}(P)$ for some $i$, $1\leq i\leq k$,
then $\beta$ has either $i$ or $i+1$
active sites.
\end{enumerate}
\end{lemma}
\proof
\comm{
For the first point, since $\alpha$ avoids $312$ and
$\alpha(i)=n$ is the largest value of the suffix 
$\alpha(i)\alpha(i+1)\ldots\alpha(n)$, it follows that this suffix is decreasing.

\noindent
For the second point, the insertion of $(n+2)$ into the $i$th site of $\beta=\alpha^{\downarrow i}$ produces no patterns in $P$ (the entries $(n+1)$ and $(n+2)$ are consecutive in $\beta$),
and the insertion of $(n+2)$ into the $(i+2)$nd site of $\beta$ produces
the forbidden pattern $312$, and since $P$ is a right-justified set of forbidden patterns the result
follows.
}
\endproof

The following result is a direct consequence of the second point of the previous lemma, 
and we state it in the next corollary in order to refer to it later.

\begin{corollary}\label{Cor}
Let $P$ be the set of forbidden patterns in Theorem \ref{T:genpad}. 
If $\alpha\in S_n(P)$, $n\geq 1$, has $k$ active sites, then $\alpha=\lambda^{\downarrow k}$
or $\alpha=\lambda^{\downarrow (k-1)}$ for some $\lambda\in S_{n-1}(P)$.
\end{corollary}

As one can see below, the pattern $231$ is of particular interest for the definition of the color of a permutation in 
Theorem \ref{T:genpad}.
If in an arbitrary permutation $\alpha\in S_n$, $n$ is involved in an occurrence of the 
pattern $231$, then $n$ plays the role of $3$ in this occurrence and we have the following easy to understand result.


To each permutation $\alpha$ in $S_n(P)$, $P=\{312,2431,(p+1)p\ldots 21\}$, we associate an integer 
$d(\alpha)\in \{0,1\}$ as: $d(\alpha)$ is $0$ if and only if at least
one of the following two conditions is fulfilled:
$n$ is not involved in an occurrence of the pattern $231$ in $\alpha$, or the length 
$p$ suffix of $\alpha$ is decreasing. Equivalently, $d(\alpha)$ is $1$ if and only if
$n$ plays the role of $3$ in an occurrence of the pattern $231$ in $\alpha$ and the length
$p$ suffix of $\alpha$ is not decreasing. Thus $d$ is a function 
$d:S(P)\to \{0,1\}$.

\proof[Proof of Theorem \ref{T:genpad}]
\comm{
Actually, we will prove by induction on $n$ the following:

\noindent
(1) the color of a permutation $\alpha\in S(P)$ defined in 
Theorem \ref{T:genpad} by means of $\nu_P$ is $ d(\alpha)$; 

\noindent
(2) the statement of Theorem \ref{T:genpad}.
\medskip

\noindent
If $n=0$, then (1) and (2) trivially hold.
\medskip

\noindent
Proof of (1):  Supposing that (1) and (2) are satisfied by length $n$ permutations,
$n\geq 0$, we prove (1) for length $n+1$ permutations.

\noindent
Let $\alpha$ be a permutation in $S_n(P)$ with $k$ active sites, $n\geq 0$.

\noindent
First we show
that  $i=1$, or $i=k$ and $d(\alpha)=0$ implies $d(\alpha^{\downarrow i})=0$.

\medskip
\noindent
If $i=1$, clearly $(n+1)$ is not involved in an occurrence of $231$ in $\alpha^{\downarrow 1}$, so $d(\alpha^{\downarrow 1})=0$.

\medskip
\noindent
If $i=k>1$ and $d(\alpha)=0$, we have $n>0$, and we distinguish two cases.
 \begin{itemize}
 \item $k=p$. By Lemma \ref{L:act1}.1, the length $p$ suffix of $\alpha^{\downarrow i}=\alpha^{\downarrow p}$
       is decreasing and so $d(\alpha^{\downarrow i})=0$.
 \item $k<p$. By Corollary \ref{Cor} it follows that  
      $\alpha=\lambda^{\downarrow (k-1)}$ or $\alpha=\lambda^{\downarrow k}$ for some 
      $\lambda\in S_{n-1}(P)$. 
      \noindent 
      If $\alpha=\lambda^{\downarrow (k-1)}$, since $\alpha$
      has $k$ active sites and $(n+1)$ and $n$ are conscutive entries in 
      $\alpha^{\downarrow k}=(\lambda^{\downarrow (k-1)})^  {\downarrow k}$
      it follows that $n$ is not involved in an occurrence of $231$
      in $\alpha$ (otherwise $\alpha^{\downarrow k}$ contains the forbidden pattern $2431$),
      and $(n+1)$ is not involved in an occurrence of $231$ in  $\alpha^{\downarrow k}$,
      so $d(\alpha^{\downarrow k})=0$.    
      \noindent 
      But $\alpha=\lambda^{\downarrow k}$ is not possible, indeed $\alpha=\lambda^{\downarrow k}$ implies either: 
      \begin{itemize}
      \item
      $\lambda$ has $j$ active sites with $j>k$,
      and in this case $n$ is involved in an occurrence of $231$ in $\alpha$ which is in contradiction 
      with $d(\alpha)=0$; or 
      \item
      $\lambda$ has (as $\alpha$) $k$ active sites which  implies that 
      the number of active sites of $\alpha$ follows the second rule in the definition of $\mu_P$
      in the statement of the present theorem,
      which in our case happens when the the color of $\lambda$ is $1$, hence the
      color of $\alpha=\lambda^{\downarrow k}$ follows the second rule in the definition of $\nu_P$,
      which again leads to a contradiction.
      \end{itemize}
\end{itemize}
\noindent
Conversely, it is routine to check that when $i>1$ and ($i<k$ or $d(\alpha)=1$) the insertion of 
$(n+1)$ into the $i$th active site of $\alpha$ produces a new occurrence of the pattern $231$, 
but not a length $p$ decreasing suffix, and so 
$d(\alpha^{\downarrow i})=1$.

\medskip
\noindent
Proof of (2):  Supposing that (1) and (2) are satisfied by length $n$ permutations,
$n\geq 0$, we prove (2) for length $n+1$ permutations.
In light of (1) it is enough to show that the number of active sites
of a permutation in $S_{n+1}(P)$ is that specified by $\mu_p$
and considering its color given by the function $d$.

Let $\alpha$ be a permutation in $S_n(P)$ with $k$ active sites and color $c$, and let
$\beta=\alpha^{\downarrow i}$ be the permutation obtained from $\alpha$ by inserting $(n+1)$ into its
$i$th active site, $1\leq i\leq k$. By Lemma \ref{L:act1}.\ref{second_p}, $\beta$ has $i$ or $i+1$ active sites.
If the condition `$i=1$ or ($i=k$ and $c=0$ and $k<p$)' is satisfied, then the insertion of 
$(n+2)$ into the $(i+1)$st site of $\beta$ produces no forbidden pattern, and so $\beta$ has
$i+1$ active sites.

By contrast, if the above condition is violated, then the insertion of 
$(n+2)$ into the $(i+1)$st site of $\beta$ produces a forbidden pattern.
To prove this, it is enough to show that $\beta^{\downarrow (i+1)}$ contains a forbidden 
pattern if $i$ satisfies one of the following conditions: 
{\it (i)} $i>1$ and ($i<k$ or $c=1$), or
{\it (ii)} $i=k=p$.

\medskip

\noindent
If $i>1$ and ($i<k$ or $c=1$), then $(n+1)$ plays the role of $3$ in an occurrence 
of the  pattern $231$ in $\alpha^{\downarrow i}$
(see the last part of the proof of (1)), and thus 
$\beta^{\downarrow (i+1)}$ contains the forbidden pattern $2431$.

\noindent
If  $i=k=p$, then $\beta^{\downarrow (i+1)}$ contains the forbidden pattern $(p+1)p\dots 1$.
}
\endproof

\begin{figure}[h]
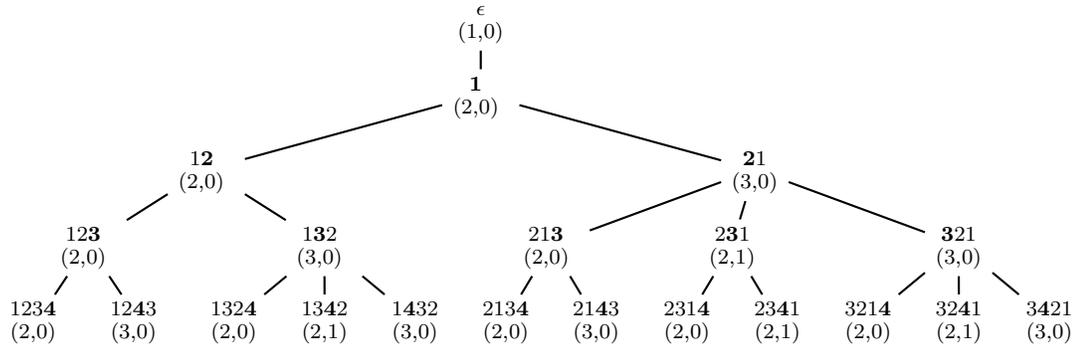

\centering    

\psset{levelsep=10mm,treesep=5mm,radius=-1mm}
\pstree[nodesep=1mm,treemode=B]{\TR{$\epsilon\atop (1,0)$}}
  {\pstree{\TR{$\mathbf{1}\atop(2,0)$ }}
    {   \pstree{\TR{$1\mathbf{2}\atop(2,0)$ }}
         { \pstree{\TR{$12\mathbf{3}\atop(2,0)$  }}
           {\pstree{\TR{$123\mathbf{4}\atop(2,0)$ }}{}
           \pstree{\TR{$12\mathbf{4}3\atop(3,0)$ }}{}
           }
          \pstree{\TR{$1\mathbf{3}2\atop(3,0)$ }}
          {\pstree{\TR{$132\mathbf{4}\atop(2,0)$}}{}
           \pstree{\TR{$13\mathbf{4}2\atop(2,1)$}}{}
           \pstree{\TR{$1\mathbf{4}32\atop(3,0)$}}{}
          }
         }
        \pstree{\TR{$\mathbf{2}1\atop(3,0)$}}
        {\pstree{\TR{$21\mathbf{3}\atop(2,0)$ }}
         {\pstree{\TR{$213\mathbf{4}\atop(2,0)$}}{}
          \pstree{\TR{$21\mathbf{4}3\atop(3,0)$}}{}
         }
         \pstree{\TR{$2\mathbf{3}1\atop(2,1)$}}
          {\pstree{\TR{$231\mathbf{4}\atop(2,0)$}}{}
           \pstree{\TR{$23\mathbf{4}1\atop(2,1)$}}{}
          }
         \pstree{\TR{$\mathbf{3}21\atop(3,0)$}}
          {\pstree{\TR{$321\mathbf{4}\atop(2,0)$}}{}
           \pstree{\TR{$32\mathbf{4}1\atop(2,1)$}}{}
           \pstree{\TR{$3\mathbf{4}21\atop(3,0)$}}{}
          }
        } 
    }
  }

    \caption{\small
The first levels of the tree induced by the succession function in Theorem \ref{T:genpad}
when $p=3$ (Corollary \ref{T:padovan}). Each node of the tree is labeled $\alpha \atop (k,c)$, with $\alpha$ 
a permutation in $S_n(P)$, $0\leq n\leq 4$, $P=\{312, 2431, 4321\}$,
$k$ the number of active sites and $c$ the color of $\alpha$. \label{fig3:chi}}
\end{figure}

In particular, when $p=3$ the set of forbidden patterns in Theorem \ref{T:genpad} 
becomes $P=\{312,2431,4321\}$ and the corresponding counting sequence is the binomial transform of Padovan sequence (\seqnum{A034943} in OEIS \cite{OEIS}). 
An illustration of the underlying tree of the succession function for this set of forbidden patterns 
is given in Figure \ref{fig3:chi}.
It turns out that in this particular case the number of active sites of a permutation in $S_n(P)$ is either 2 or 3 (except for the empty permutation $\epsilon$),
and we have the next corollary.
\begin{corollary} \label{T:padovan}
The succession function for the set of forbidden patterns 
$P=\{312,2431,4321\}$ is given by:
\begin{itemize}
\item[]
$\mu_P(i,k,c)=\left\{\begin{array}{ll}
            2 & \mbox{if $i=1$ or $i=2$ and ($c=1$ or $k=3$)} \\
            3 & \mbox{otherwise,}
            \end{array}   \right. $
\end{itemize}
\noindent
and
\begin{itemize}
\item[]
$\nu_P(i,k,c)=\left\{\begin{array}{ll}
            0 & \mbox{if $i=1$ or ($i=k$ and $c=0$)} \\
            1 & \mbox{otherwise.}
            \end{array}   \right.$
\end{itemize}
    \end{corollary}

Notice that the binomial transform of Padovan sequence also counts permutations avoiding $\{321,2413,3142\}$, see \cite{Vat12}. However, by Theorem \ref{P:rjp2}, this set of forbidden patterns is not right-justified whereas $P=\{312,2431,4321\}$ is regular.  

In the same vein, we give below without proof the colored succession function for the second 
set of patterns in Example \ref{ex_d}.\ref{bisection}. Its proof is similar with that of 
Theorem \ref{T:genpad} when the variable length pattern is omitted.
 
\begin{proposition}\label{Th.34-Fib}
The succession function for the set of forbidden patterns $P=\{312,2431\}$ 
is $\chi_P(i,k,c)=(\mu_P(i,k,c),\nu_P(i,k,c))$, with
\begin{itemize}
\item[]
$\mu_P(i,k,c)=\left\{\begin{array}{ll}
            i+1 & \mbox{if $i=1$ or ($i=k$ and $c=0$)} \\
            i & \mbox{otherwise,}
            \end{array}   \right. $
\end{itemize}
\noindent
and
\begin{itemize}
\item[]
$\nu_P(i,k,c)=\left\{\begin{array}{ll}
            0 & \mbox{if $i=1$ or ($i=k$ and $c=0$)} \\
            1 & \mbox{otherwise.}
            \end{array}   \right.$
\end{itemize}
    \end{proposition}

We give succession functions for some sets of c-regular forbidden patterns in Table \ref{table3} in Appendix.

\section{Efficient generating algorithms} \label{sec-CAT}
In this section, we present exhaustive generating algorithms
for permutations avoiding a set of forbidden regular and c-regular patterns and we show 
that they are efficient.
 
For a set of forbidden patterns $P$ and for $\alpha\in S_n(P)$,
since $\alpha^{\downarrow (i+1)}=(\alpha^{\downarrow i})^{\leftarrow}$, we have that 
$\alpha^{\downarrow i}$ and $\alpha^{\downarrow (i+1)}$ differ by a transposition
of two adjacent entries, one of them being $(n+1)$.
In our algorithms we represent permutations  $\alpha\in S_n$ by length $n$ global arrays.
The insertion of an element into an array is not an efficient 
operation, except when a next largest entry is inserted after the last entry of the array.
By contrast, the transposition of two adjacent elements requires only a constant time and
we express insertions by iterating transpositions of adjacent entries, and this is crucial for the efficiency of our generating
algorithms.
In the algorithms in Table \ref{Alg1-2}, $\alpha \cdot (i,j)$ is the permutation obtained from $\alpha$
by transposing the entries in positions $i$ and $j$; and $[\alpha, length]$ is the 
permutation in $S_{length}(P)$ obtained from $\alpha\in S_{length-1}(P)$ by inserting the integer 
$length$ after the last entry of $\alpha$, that is the permutation $\alpha^{\downarrow 1}$.

Given an explicit implementation of the succession function $\chi_P$,
the generating algorithms in Table \ref{Alg1-2} produce exhaustively the length $n$ 
$P$-avoiding permutations, for a set of regular or c-regular forbidden patterns $P$.
In particular, the algorithm (a) (which is first mentioned by Duckes et al. \cite{DFMV08} in the context of Gray code generation) mimes the succession rules given after
Definition \ref{def_reg_p};
and tree induced by the recursive calls of the algorithm (b) for the permutations in
Corollary \ref{T:padovan} with $n\leq 4$ is depicted in Figure \ref{fig3:chi}.

\begin{table}[h] 
\begin{tabular}{cc}
 \fbox{\small
\begin{minipage}{0.47\textwidth}
\begin{algorithmic}
\STATE {\bf procedure} Gen\_Avoid($length,k$) \STATE {\bf local} $i$
\IF{$length=n$} \STATE Print($\alpha$) \ELSE
      \STATE $length:=length+1$
      \STATE $\alpha:=[\alpha, length]$
      \STATE Gen\_Avoid($length,\chi_P(1,k)$)
      \FOR{$i:=2$ {\bf to} $k$}
      \STATE $\alpha:=\alpha \cdot (length-i+2,length-i+1)$
      \STATE Gen\_Avoid($length,\chi_P(i,k)$)
      \ENDFOR
      \FOR{$i:=k$ {\bf downto} $2$}
      \STATE $\alpha:=\alpha \cdot (length-i+2,length-i+1)$
      \ENDFOR
\ENDIF \STATE {\bf end procedure}
\end{algorithmic}
\end{minipage}
}
\hspace{0.1cm}
&
 \fbox{\small
\begin{minipage}{0.47\textwidth}
\begin{algorithmic}
\STATE {\bf procedure} Gen\_Avoid($length,k,c$)

\STATE {\bf local} $i,u,v$ 
      \IF{$length=n$} 
      \STATE Print($\alpha$) \ELSE
      \STATE $length:=length+1$
      \STATE $\alpha:=[\alpha, length]$
      \STATE $(u,v):=\chi_P(1,k,c)$
      \STATE Gen\_Avoid($length,u,v$)
      \FOR{$i:=2$ {\bf to} $k$}
      \STATE $\alpha:=\alpha \cdot (length-i+2,length-i+1)$
      \STATE $(u,v):=\chi_P(i,k,c)$
      \STATE Gen\_Avoid($length,u,v$)
      \ENDFOR
      \FOR{$i:=k$ {\bf downto} $2$}
      \STATE $\alpha:=\alpha \cdot (length-i+2,length-i+1)$
      \ENDFOR
\ENDIF \STATE {\bf end procedure}
\end{algorithmic}
\end{minipage}
}
\\
(a)&(b)\\
\end{tabular}
\caption{\small (a) Algorithm for generating permutations avoiding:
(a) a regular pattern characterized by the succession function $\chi_P$, with the initial call
Gen\_Avoid($0,1$); and  
(b) a c-regular pattern characterized by the succession function
$\chi_P$, with the initial call Gen\_Avoid($0,1,0$). 
In both cases the initial permutation is the length zero permutation $\epsilon$.
\label{Alg1-2}} 
\end{table}

A recursive generating algorithm is said to run in {\it constant amortized time} (CAT) if it generates each object in $O(1)$ time, in amortized sense. Such an algorithm is also called a {\it CAT algorithm}. 
The following {\it CATness} principle is due to Frank Ruskey.

\begin{proposition}{(\cite{Rus17})}\label{catness}
A recursive generating algorithm is a CAT one if it satisfies the following properties:
\begin{itemize}
\item Each recursive call generates at least one object (there is no dead-end recursive call);
\item The amount of computation in each recursive call is proportional to the degree of the call (that is, the number of subsequent recursive calls produced by the current call);
\item The number of recursive calls having degree one (if any) is $O(N)$, where $N$ is the number of generated objects.
\end{itemize}
\end{proposition}

Let $P$ be a set of regular or c-regular forbidden patterns, and $\alpha\in S_n(P)$, $n\geq 0$.
By the defintion of regularity, $\alpha ^{\downarrow 1}$ belongs to $S_{n+1}(P)$, and if
$\alpha ^{\downarrow 1}$ has only one active site (or equivalenty, 
$(\alpha^{\downarrow 1})^{\downarrow 2}$
contains a pattern in $P$), then there is a length $k\geq 2$ permutation 
$\tau=\tau(1)\tau(2)\ldots \tau(k-2)k(k-1)$ belonging to $P$.

The number of recursive calls produced by a current call of our algorithms is given by $\chi_P$, and
combining Proposition \ref{catness} with the considerations above we have the following theorem.

\begin{theorem}\label{eff}
If $P$ is a set of regular or c-regular forbidden patterns for which the corresponding 
succession function $\chi_P$ can be computed in constant time, and $P$ 
does not contain patterns $\tau$ of the form $\tau(1)\tau(2)\ldots \tau(k-2)k(k-1)$, $k\geq 2$, then the algorithms in Table \ref{Alg1-2}
generate in constant amortized time the set $S_n(P)$, $n\geq 0$.
\end{theorem}

In Tables \ref{table2} and \ref{table3} in Appendix we list several regular and c-regular
sets of forbidden patterns satisfying Theorem \ref{eff}.
For some of them, the corresponding succession functions are given 
in Sections \ref{S:3} and \ref{S:4} of the present paper.

The generating order of our algorithms is not the lexicographical one, and we have the next proposition.

\begin{proposition}
Let $L_n(P)$, $n\geq 0$, be the ordered list for the set $S_n(P)$ produced by algorithms
in Table \ref{Alg1-2}.
Then $\alpha$ precedes $\beta$ in this list if either $n\geq 1$ and $\alpha'$ precedes $\beta'$ in 
$L_{n-1}(P)$, where $\alpha'$ and $\beta'$ are the permutations obtained from
$\alpha$ and $\beta$ by erasing their largest element $n$; or $i<j$, where $i$ and $j$
are the positions (from right to left) of $n$ in $\alpha$ and $\beta$.

\end{proposition}

Finally, if for a set of right-justified forbidden patterns $P$, $P^r$ (resp. $P^c$) denotes the set
of patterns obtained by reversing (resp. complementing) each pattern in $P$ (see for example \cite{Kitaev2011}
for the definition of these two operations), then our algorithms can easily be adapted to generate
$S_n(P^r)$ and $S_n(P^c)$ provided they generate $S_n(P)$.

\section*{Acknowledgment}
This research is funded by Vietnam National Foundation for Science and Technology Development (NAFOSTED) under grant number 102.01-2016.05. A part of this manuscript was accomplished when the second author was visiting the Vietnam Institute for Advanced Study in Mathematics (VIASM).


\bibliographystyle{plain}


\newpage
\section*{Appendix}
\begin{table} [h] 
{\scriptsize
\[\begin{array}{|l|l|l|} \hline
\mbox{\hspace{1cm} Counting sequence/class}           & \hspace{1cm} P  & \hspace{1cm}  \chi_P(i,k) \\
\hline 
\hline
\begin{array}{l}
    \\
    \mbox{$2^{n-1}$\cite{BBP06}}\\
\end{array}
                        & \{321,312\}     & \begin{array}{ll}
                                                2
                                            \end{array} \\ 
\hline
\begin{array}{l}
    \\
    \mbox{Pell numbers \cite{BBP06,Gui95}}\\
    \mbox{(\seqnum{A000129} in OEIS \cite{OEIS})}
\end{array}
                        & \begin{array}{l}\hspace{-2mm}\{321,3412,4123\}\\ 
                                          \mbox{particular instance of Theorem \ref{Th.3pm}}
                          \end{array}
                                            & \begin{array}{ll}
                                                3 & \mbox{if $i = 1$} \\
                                                2 & \mbox{otherwise.}
                                              \end{array} \\ \cline{2-3}
                        & \{312,4321,3421\}
                                            & \begin{array}{ll}
                                                3 & \mbox{if $i = 2$} \\
                                                2 & \mbox{otherwise.}
                                            \end{array} \\
\hline

                        & \begin{array}{l}\hspace{-2mm}\{321,3412\}\\ 
                                          \mbox{particular instance of Theorem \ref{Th.3pm}}
                          \end{array}

                                            & \begin{array}{ll}
                                                k+1 & \mbox{if $i = 1$} \\
                                                2 & \mbox{otherwise.}
                                              \end{array} \\ \cline{2-3}

\begin{array}{l}
    \mbox{Bisection of Fibonacci sequence \cite{BBP06,Gui95}} \\
    \mbox{(\seqnum{A001519} in OEIS \cite{OEIS})}
\end{array}
                        & \begin{array}{l}\hspace{-2mm} \{321,4123\}\\ 
                                          \mbox{particular instance of Theorem \ref{Th.3pm}}
                          \end{array}
                                            & \begin{array}{ll}
                                                3 & \mbox{if $i = 1$} \\
                                                i & \mbox{otherwise.}
                                              \end{array} \\ \cline{2-3}
                        & \{312,4321\}
                                            & \begin{array}{ll}
                                                3 & \mbox{if $k = 3$ and $i=3$} \\
                                                i+1 & \mbox{otherwise.}
                                              \end{array} \\
\hline

\begin{array}{l}
    \mbox{Catalan numbers \cite{Wes94}}\\
    \mbox{(\seqnum{A000108} in OEIS \cite{OEIS})}
\end{array}
                        & \{312\}
                                            & \begin{array}{ll}
                                                i+1
                                              \end{array} \\ \cline{2-3}
                        & \{321\}
                                            & \begin{array}{ll}
                                                k+1 & \mbox{if $i = 1$} \\
                                                i & \mbox{otherwise.}
                                            \end{array} \\
\hline

\begin{array}{l}
    \mbox{Schr\"oder numbers \cite{Gui95}} \\
    \mbox{(\seqnum{A006318} in OEIS \cite{OEIS})}
\end{array}
                        & \{4123,4213\}
                                            & \begin{array}{ll}
                                                k+1 & \mbox{if $i = k-1$ or $i=k$} \\
                                                i+2 & \mbox{otherwise.}
                                              \end{array} \\
\hline

\begin{array}{l}
    \mbox{Fibonacci numbers \cite{BD06}}\\
    \mbox{(\seqnum{A000045} in OEIS \cite{OEIS})}
\end{array}
                        & \begin{array}{l}\hspace{-2mm}\{321,231,312\}\\ 
                                          \mbox{particular instance of Theorem \ref{Th.3pm}}
                          \end{array}
                                            & \begin{array}{ll}
                                                1 & \mbox{if $i = 2$} \\
                                                2 & \mbox{otherwise.}
                                              \end{array} \\
\hline

                        & \begin{array}{l}\hspace{-2mm}\mbox{$\{321,(p+1)12\ldots p\}$ \cite{BBP06,CW99}}\\ 
                                          \mbox{particular instance of Theorem \ref{Th.3pm}}
                          \end{array}
                        
                                            & \begin{array}{ll}
                                                k+1 & \mbox{if $i = 1$ and $k< p$} \\
                                                p & \mbox{if $i = 1$ and $k = p$} \\
                                                i & \mbox{otherwise.}
                                              \end{array} \\ \cline{2-3}
\begin{array}{l}
    \mbox{A pattern of length 3 and} \\
    \hspace{5mm} \mbox{a variable length pattern} \\
    \\
\end{array}
                        & \begin{array}{l}\hspace{-2mm}\mbox{$\{321,p(p+1)12\ldots (p-1)\}$ \cite{BBP06,CW99}}\\ 
                                          \mbox{particular instance of Theorem \ref{Th.3pm}}
                          \end{array}
                        
                                            & \begin{array}{ll}
                                                k+1 & \mbox{if $i = 1$} \\
                                                i & \mbox{if $1<i<p-1$} \\
                                                p-1 & \mbox{otherwise.}
                                              \end{array} \\ \cline{2-3}
                        & \mbox{$\{312,(p+1)p\ldots 21\}$ \cite{CW99}}
                                            & \begin{array}{ll}
                                                p & \mbox{if $k = p$ and $i = p$} \\
                                                i+1 & \mbox{otherwise.}
                                              \end{array} \\
\hline
\begin{array}{l}
    \mbox{A pattern of length 3,} \\
    \hspace{2mm} \mbox{a pattern of length 4 and} \\
    \hspace{2mm} \mbox{a variable length pattern}
\end{array}
                        & \begin{array}{l}\hspace{-2mm}\mbox{$\{321,3412,(p+1)12\ldots p\}$ \cite{BBP06}}\\
                            \mbox{particular instance of Theorem \ref{Th.3pm}}
                          \end{array}
                                            & \begin{array}{ll}
                                                k+1 & \mbox{if $i = 1$ and $k < p$} \\
                                                p & \mbox{if $i = 1$ and $k = p$} \\
                                                2 & \mbox{otherwise.}
                                              \end{array} \\
\hline

\begin{array}{l}
    \mbox{Generalized Fibonacci } \\
    \hspace{2mm} \mbox{numbers} 
\end{array}
                        & \begin{array}{l}\hspace{-2mm}\mbox{$\{321,231,(p+1)12\ldots p\}$ \cite{BBP06}}\\
                                          \mbox{particular instance of Theorem \ref{Th.3pm}}
                          \end{array}
                                            & \begin{array}{ll}
                                                k+1 & \mbox{if $i=1$ and $k<p$} \\
                                                k & \mbox{if $i=1$ and $k=p$} \\
                                                1 & \mbox{otherwise.}
                                              \end{array} \\
\hline

\begin{array}{l}
    \\ \\
    \mbox{A pattern of length 3 and} \\
    \hspace{2mm} \mbox{two variable length patterns}\\
\end{array}
                        & \begin{array}{l}\hspace{-2mm}\mbox{$\{321,p(p+1)12\ldots (p-1),(p+1)12\ldots p\}$ \cite{BBP06}}\\
                                          \mbox{particular instance of Theorem \ref{Th.3pm}}
                          \end{array}
                                        
                                            & \begin{array}{ll}
                                                k+1 & \mbox{if $i = 1$ and $k < p$} \\
                                                p & \mbox{if $i = 1$ and $k = p$} \\
                                                p-1 & \mbox{if $i = p$ and $k = p$} \\
                                                i & \mbox{otherwise.}
                                              \end{array} \\ \cline{2-3}
                        & \begin{array}{l}\mbox{\hspace{-2mm}$\{321,p(p+1)12\ldots (p-1),(m+1)12\ldots m\}$}\\
                           \hspace{2mm} \mbox{Theorem \ref{Th.3pm}}
                           \end{array}
                                            & \begin{array}{ll}
                                                k+1 & \mbox{if $i = 1$ and $k<m$} \\
                                                m & \mbox{if $i = 1$ and $k = m$} \\
                                                i & \mbox{if $1<i<p$} \\
                                                p-1 & \mbox{otherwise.}
                                              \end{array} \\
\hline

\end{array}\]
}
\caption{\small A sample of regular forbidden patterns $P$.
Their succession function (last column) is either folklore or easy to check,
or given in the corresponding reference. \label{table2}}
\end{table}



\begin{table} [h] 
{\scriptsize
\[\begin{array}{|l|l|l|l|} \hline
\mbox{\hspace{1cm} Counting sequence/class}  & \hspace{1cm} { P}  & \hspace{1cm} { \mu_P(i,k,c)}  & \hspace{1cm} { \nu_P(i,k,c)}\\
\hline \hline
\begin{array}{l}
    \mbox{Bisection of Fibonacci sequence} \\
    \mbox{(\seqnum{A001519} in OEIS \cite{OEIS})}
\end{array}
                        & \begin{array}{l}\hspace{-2mm}\{312,2431\}\\
                           \mbox{Proposition } \ref{Th.34-Fib}\end{array}
                                            & \begin{array}{ll}
                                                i+1 & \mbox{if $i=1$ or}\\
                                                &\hspace{2mm}\mbox{($i=k$ and $c=0$)} \\
                                                i & \mbox{otherwise.}
                                              \end{array}
                                                                    & \begin{array}{ll}
                                                                        0 & \mbox{if $i=1$ or}\\
                                                                        & \hspace{2mm}\mbox{($i=k$ and $c=0$)} \\
                                                                        1 & \mbox{otherwise.}
                                                                      \end{array}   \\
\hline
\begin{array}{l}
    \mbox{Binomial transform of} \\
    \hspace{2mm} \mbox{Padovan sequence} \\
    \hspace{2mm} \mbox{(\seqnum{A034943} in OEIS \cite{OEIS})}
\end{array}
                        & \begin{array}{l}\hspace{-2mm}\{312,2431,4321\}\\
                                         \mbox{Corollary \ref{T:padovan}}
                                         \end{array}
                                            & \begin{array}{ll}
                                                2 & \mbox{if $i=1$ or $i=2$ and}\\
                                                   &\hspace{2mm}\mbox{($c=1$ or $k=3$)} \\
                                                3 & \mbox{otherwise.}
                                              \end{array}
                                                                    & \begin{array}{ll}
                                                                        0 & \mbox{if $i=1$ or} \\
                                                                        & \hspace{2mm}\mbox{($i=k$ and} \\
                                                                        & \hspace{2mm}\mbox{$c=0$)} \\
                                                                        1 & \mbox{otherwise.}
                                                                      \end{array}   \\
\hline

\begin{array}{l}
    \mbox{Generalized Fibonacci}\\
    \hspace{2mm} \mbox{numbers}
\end{array}
                        & \mbox{$\{321,312,23\ldots (p+1)1\}$ \cite{BD06}}
                                            & \begin{array}{ll}
                                                1 & \mbox{if $c=p-2$} \\
                                                &\hspace{2mm}\mbox{and $i=2$} \\
                                                2 & \mbox{otherwise.}
                                              \end{array}
                                                                    & \begin{array}{ll}
                                                                        0 & \mbox{if $i=1$ or  } \\
                                                                        &\hspace{2mm}\mbox{$c=p-2$} \\
                                                                        c-1 & \mbox{otherwise.}
                                                                      \end{array}   \\
\hline 
\begin{array}{l}
    \mbox{A 3-length pattern,}\\
    \hspace{2mm} \mbox{a 4-length pattern}\\
    \hspace{2mm} \mbox{and a variable} \\
    \hspace{2mm} \mbox{length pattern}
\end{array}
                        & \mbox{$\{321,4123,34\ldots (p+1)12\}$ \cite{BBP06}}
                                            & \begin{array}{ll}
                                                2 & \mbox{if $i=2$ or ($i=3$} \\
                                                &\hspace{2mm}\mbox{ and $c=p-3$)} \\
                                                3 & \mbox{otherwise.}
                                              \end{array}
                                                                    & \begin{array}{ll}
                                                                        c+1 & \mbox{if $i=3$} \\
                                                                        &\hspace{2mm}\mbox{and $c<p-3$} \\
                                                                        0 & \mbox{otherwise.}
                                                                      \end{array}   \\ \cline{2-4}

                        & \begin{array}{l}\mbox{\hspace{-2mm}$\{312,2431,(p+1)p\ldots 21\}$}\\
                          \mbox{Theorem }\ref{T:genpad}\end{array}
                                            & \begin{array}{ll}
                                                i+1 & \mbox{if $i=1$ or} \\
                                                &\hspace{2mm}\mbox{($i=k<p$ and} \\
                                                &\hspace{2mm}\mbox{$c=0$)} \\
                                                i & \mbox{otherwise.}
                                              \end{array}
                                                                    & \begin{array}{ll}
                                                                        0 & \mbox{if $i=1$ or} \\
                                                                          &\hspace{2mm}\mbox{($i=k<p$ and} \\
                                                                          &\hspace{2mm}\mbox{$c=0$)} \\
                                                                        1 & \mbox{otherwise.}
                                                                      \end{array}   \\ \cline{2-4}
\hline

\end{array}\]
}
\caption{A sample of c-regular forbidden patterns $P$ together with the
$\chi_P(i,k,c)=(\mu_P(i,k,c),\nu_P(i,k,c))$ functions. \label{table3}}
\end{table}

\end{document}